# Single cell resolution 3D imaging and segmentation within intact live tissues

G. Paci[1,2,*,#], P. Vicente-Munuera[1,2,*,#], I. Fernandez-Mosquera[1,2], A. Miranda[1,2], K. Lau[1,2], Q. Zhang[1,2], R. Barrientos[1,2], Y. Mao[1,2,*,#]

1. Laboratory for Molecular Cell Biology, University College London, London, UK.
2. Institute for the Physics of Living Systems, University College London, London, UK
* These authors contributed equally
# E-mail to: g.paci@ucl.ac.uk, p.munuera@ucl.ac.uk, y.mao@ucl.ac.uk

## Abstract

Epithelial cells form diverse structures from squamous spherical organoids to densely packed pseudostratified folded tissues. Quantification of cellular properties in these contexts requires high-resolution deep imaging and computational techniques to achieve truthful three-dimensional (3D) structural features. Here, we describe a detailed step-by-step protocol for sample preparation, imaging and deep-learning-assisted cell segmentation to achieve accurate quantification of fluorescently labelled individual cells in 3D within live tissues. We share the "lessons learned" through troubleshooting 3D imaging of *Drosophila* wing discs, including considerations on the choice of microscopy modality and settings (objective, sample mounting) and available segmentation methods. In addition, we include a computational pipeline alongside custom code to assist replication of the protocol. While we focus on the segmentation of cell outlines from membrane labelling, this protocol applies to a wide variety of samples, and we believe it will be valuable for studying other tissues that demand complex analysis in 3D.

## Introduction

Epithelial cells can display an astounding variety of cell shapes, from very thin and flat cells in squamous epithelia to extremely tall and tortuous in densely packed pseudostratified epithelia. Technological development in microscopy including 2- and 3-photon excitation, adaptive optics and improved objectives now allow imaging these tissues at higher resolution and depth (1, 2), unlocking the possibility of single-cell resolution quantification. Artificial Intelligence (AI) is in vogue, and its impact on cell biology is equally significant (3, 4). The use of AI algorithms aims at reducing the time for human annotations and corrections. In particular, the process of identifying individual cells (semantic segmentation) in 3D can be very time consuming (5). Deep learning algorithms (a subfield within AI) has proven useful to analyse tissues in 3D (6-8). However, the accuracy of deep learning models on a given dataset relies on its similarity to the dataset used to train it. To tackle this, transfer learning has been used, where only the last layer of weights is re-trained on a pre-trained neural network. However, to utilize transfer learning, annotated ground truth data is required and generating it via manual annotation is very time-consuming. A solution to this has been proposed: human-in-the-loop (9, 10) system that combine the advantages of AI-assisted prediction with minimal expert user input, which is the inspiration for the pipeline proposed here.

Here, we present an end-to-end protocol for imaging and segmenting individual cells in an example pseudostratified epithelium, the *Drosophila* wing disc (Figure 1). Wing disc cells are densely packed (11) and have complex and highly heterogeneous 3D shapes with multiple neighbour exchanges occurring along the apico-basal axis (scutoids (12)). The high cellular density, thickness and curvature of the tissue all increase scattering and aberrations during fluorescence imaging, providing challenging conditions for individual cell segmentation. Our protocol presents steps for mounting and imaging membrane-labelled samples optimised to minimise scattering and signal degradation in the axial direction. We also present a human-in-the-loop pipeline and associated jupyter notebook (13) to perform 3D segmentation of individual cells with minimal manual corrections. This leverages on popular open-access tools including Cellpose (14), TrackMate (15) and napari (16). Where relevant, we discuss troubleshooting steps and alternatives that can be adopted where specific equipment is not available (marked by **). Importantly, we also report on widely adopted strategies that we tested but were unsuccessful in improving cell segmentation and discuss metrics to evaluate the quality of 3D cell labels. Overall, we anticipate that this protocol will be very useful to guide researchers interested in obtaining single-cell morphometric quantifications from dense tissues or particularly challenging cell shapes.

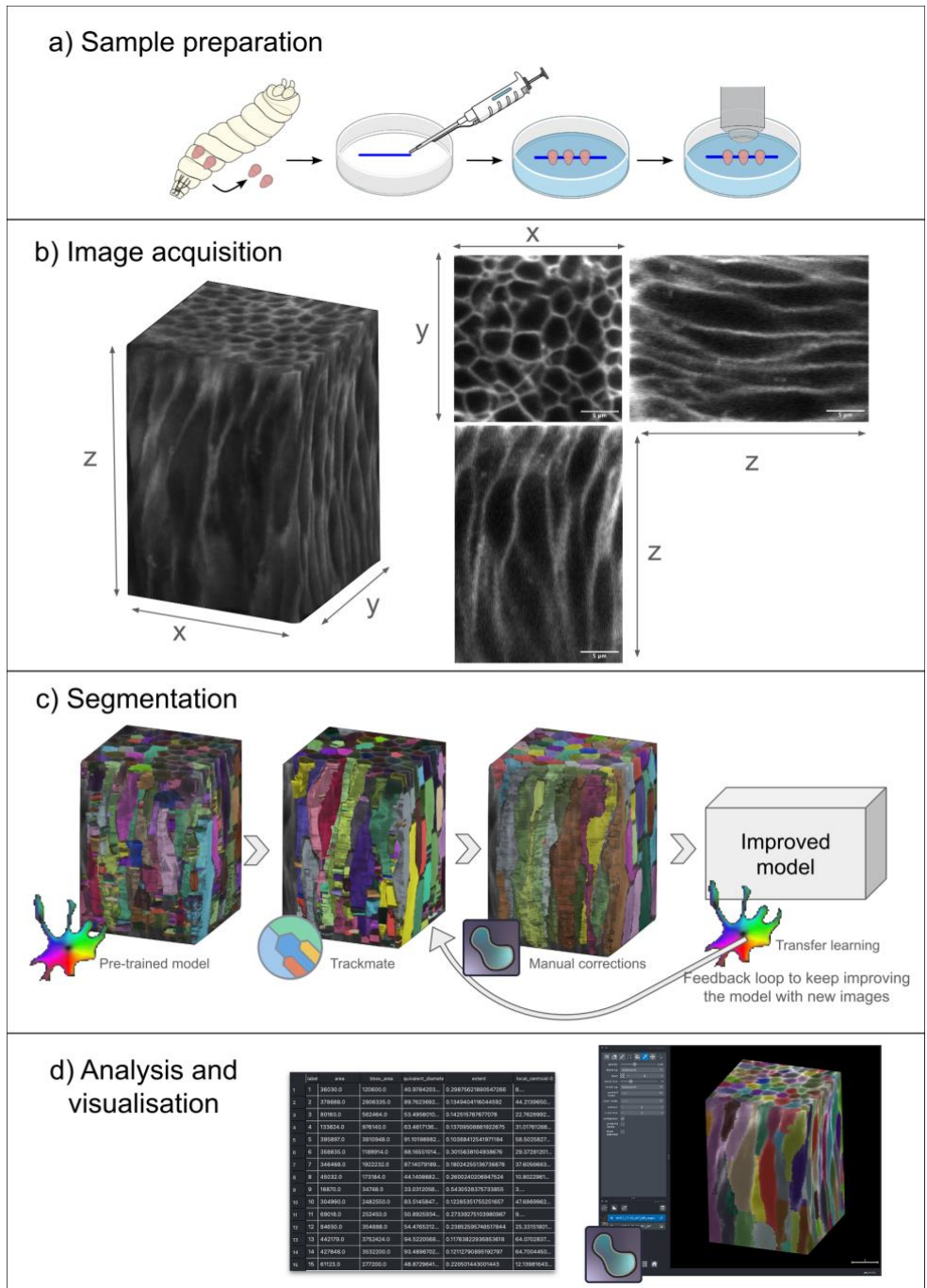

**Figure 1.** Protocol overview entailing sample preparation (a), image acquisition (b), segmentation (c), and analysis (d). a) Wing discs are dissected from 3rd instar larvae and mounted on the bottom of a plastic cell culture dish using a biocompatible adhesive to be imaged with a dipping objective in media. b) Example images acquired using the protocol described. 3D perspective (left), and orthogonal views (right) are shown. Scale bar is 5 µm. c) Segmentation pipeline, from left to right: initial segmentation with Cellpose cyto3 model, automatic stitching correction with TrackMate, manual annotation with napari, and transfer learning to obtain a new model. d) Visualization and analysis step with napari. Left: example table. Right: 3D visualization.

# Methods

## 1. Materials

### Fly stocks

*Drosophila melanogaster* membrane-labelled lines yw; Ubi-GFP-CAAX (CAAX-GFP, DGRC 109824 FBID: FBtp0011013) or NubGal4, UAS-myrGFP (nubbin-Gal4 FBID: FBti0016825 recombined with myrGFP, gift of Thompson group).

### Reagents

Corning Cell-Tak Cell and Tissue Adhesive, 1 mg (product number 354240)
Round plastic dish – 60 Thermo Scientific Nunc Cell Culture Dishes (product number 150462)
Fluorescent beads - PS-Speck Microscope Point Source Kit (Invitrogen, product number P7220)

### Equipment

#### *Microscopy*

Leica Microsystems Stellaris 8 DIVE microscope equipped with a 25x water immersion HC IRAPO L motCORR objective (NA=1) and a dual-beam Coherent Discovery multiphoton laser with a tuneable laser line for excitation.

#### *Workstations*

Two computers were used to perform the segmentation protocol: a Windows and an Ubuntu Linux machine. Both with the following hardware:
- **Base**: Dell Precision 5820 Tower XCTO.
- **Processor**: Intel® Xeon® W-2223 (8.25 MB cache, 4 cores, 8 threads, 3.60 GHz to 3.90 GHz Turbo, 120 W).
- **Graphic card**: NVIDIA® RTX™ A5000, 24 GB GDDR6, 4 DP.
- **RAM memory**: 64 GB, 4 x 16 GB, DDR4, 2933 MHz, ECC.
- **Two hard drives**: 2 TB, 7200 RPM, 3.5-inch, SATA, HDD; 512 GB, M.2, PCIe NVMe, SSD, Class 40.

#### *Software*

Hyugens Professional version (Scientific Volume Imaging, The Netherlands, http://svi.nl), section 3.1.
Cellpose3 (14, 17) https://www.cellpose.org , sections 3.1, 4.1, and 4.5.
Napari (16) https://napari.org/stable/ with the following collection of plugins installed: 'devbio-napari' https://github.com/haesleinhuepf/devbio-napari, sections 4.2, and 5; EpiTools (18, 19) https://github.com/epitools/epitools, section 4.3.
FIJI: Fiji is just ImageJ (20) https://imagej.net/software/fiji/ , sections 4.4, and 5.1.
TrackMate (15) https://imagej.net/plugins/trackmate/ , section 4.4.

## 2. *Drosophila melanogaster* husbandry, wing disc dissection and mounting

*Fly husbandry:*

1. Raise fly stocks of the desired genotype on standard cornmeal molasses fly food medium at 25°C. Per 1L, the fly food contained 10g agar, 15g sucrose, 33g glucose, 35 g years, 15g maize meal, 10g wheat germ, 30g treacle, 7.22g soya flour, 1g nipagin, 5mL propionic acid. To visualise cell membranes, we used flies of the genotype ubi-CAAX-GFP or NubGal4, UAS-MyrGFP.
2. 5-6 days before the target experiment day, flip the flies onto a new fresh food vial.

*Wing disc dissection and mounting:*

1. Remove third instar larvae gently from the food vials, rinse them in PBS and transfer them into a glass dissection well. Dissect wing imaginal discs using forceps in culture media: Shields and Sang M3 media (Merck) supplemented with 2% FBS (Merck), 1% pen/strep (Gibco), 3 ng/ml ecdysone (Merck) and 2 ng/ml insulin (Merck).
2. Prepare dishes for mounting: pipette a thin horizontal stripe of Cell-Tak at the centre of a plastic round cell culture dish and place it on a heated plate for 10 min to dry.
3. Mount wing discs: fill the dish with 5 mL of culture media, aspirate the wing discs with a 20 µL pipette and transfer them into the dish, away from the Cell-Tak. By moving the forceps rapidly in proximity to the discs (whisking motion), gently move the discs without touching them by causing local media flow. In this way, discs float slightly and can be moved to the desired position onto the Cell-Tak strip, letting them settle down.
4. Bring samples to the microscope to be imaged immediately after mounting.

**As an alternative mounting option, discs can be mounted between a coverslip and rectangular coverglass using two strips of double-sided tape to create a small vertical channel filled with 10 µL of medium. This is suitable for imaging on both upright and inverted microscopes.

## 3. Imaging

For imaging, the steps are as follows:

1. Bring the mounted dish to the upright microscope and gently dip the objective lens in the media, use brightfield imaging to identify and focus on the wing disc to be imaged.
   **In our experience, objective choice and optimal sample mounting is critical to minimise signal loss at depth: a dipping objective minimises refractive index mismatches and multiphoton laser attenuation.
2. Tune the laser line to a wavelength of 924 nm for two-photon GFP imaging. Adjust laser intensity to avoid saturation of the signal on the apical side while maximising the signal at deeper planes. Adjust zoom to the region of interest to be imaged, we found a zoom of 8 to be a good compromise.
   **We found that the fluorophore excitation peaks set by default in the microscope software are not always optimal so recommend initially testing a set of wavelengths to maximise the signal obtained.

3. Acquire stacks with a 0.5 µL spacing spanning the whole tissue (approximately 100 planes for a typical wing disc).
   \*\*For live tissue imaging, we found that it's essential to optimise imaging parameters to enable rapid acquisition of the full 3D stack (to a maximum of around 10 minutes), as dynamic cell movements may otherwise give rise to blurring motion artefacts.

### 3.1. Pre-processing: image restoration

Deconvolution is a powerful approach that aims to restore an image degraded by blurring and noise to improve image quality while preserving the structures of interest. We tested two deconvolution/deblurring strategies to see whether they could improve image quality and segmentation outcomes.

*Deconvolution with a measured PSF*

For the most accurate results, an experimental point spread function (PSF) needs to be measured using fluorescent beads on the same microscope used to acquire the images to be segmented:

1. Prepare a sample of fluorescent beads and image it using the same microscopy settings used for the samples to be segmented.
2. In Hyugens, distil a PSF using the beads image.
3. In Hyugens, deconvolve the samples to be segmented using the measured PSF.
4. Once you have obtained a deconvolved image, you can drag and drop it into Cellpose. In Cellpose, segment the image using the steps explained in the *"Initial segmentation: Cellpose"* section.

\*\* We used Hyugens software but many alternative deconvolution software are available.

*Denoising with Cellpose 3.0*
1. Denoise/deblur/upscale your images using the 'one_click3' model. Go to the GUI and press 'One_click3'.
2. Press 'run model' to obtain the segmentation.

\*\*You can find more technical instructions in the '3D-deep-segmentation-protocol' jupyter notebook (section "2. Improving raw images for segmentation: Denoising").

Overall, while both image restoration methods improved the visual appearance of the images tested, this did not correspond to improved segmentation results. In particular, running the segmentation pipeline on deconvolved images resulted in highly fragmented labels. We did however find that acquiring images of beads was a useful reference for checking the PSF quality and helped us fine-tune some microscopy parameters, for example the objective correction collar.

## 4. Segmentation

To obtain a 3D instance segmentation at a single cellular level (each cell individually segmented), we have tried different software and pipelines (see Discussion). The approach that gave best results starting with no ground truth data is the following (Figure 1c):

1) Obtain an initial segmentation with Cellpose, using the pre-trained 'cyto3' model.
2) Manually correct the segmentation of each individual 2D slice.
3) Use TrackMate to correct 3D stitching issues automatically and manually correct the remaining issues.
4) Re-train the 'cyto3' model with the corrected ground truth dataset.
5) Repeat 1-4 with the new model re-trained for every new image to be segmented.

To reproduce these steps and the following sections, we have developed a jupyter notebook: https://colab.research.google.com/drive/1ToQjW9W42gO1wZmQa4qYr4GbYbPGOS-w

**For a first attempt, we recommend using Google Colab to get an intuition on how things work. Google Colab is an on-line jupyter notebook that requires no setup and is free to use. For longer fine-tuning training sessions, Google Colab has a policy that tasks cannot be run for longer than 1 day. Thus, we recommend using a local computer for longer fine-tuning sessions.
**We observed that Google Colab may close your session if you have exceeded its (not public) usage.

### 4.1. Initial segmentation: Cellpose

1. Install Cellpose, according to your operating system, from:
   https://github.com/mouseland/cellpose?tab=readme-ov-file#option-1-installation-instructions-with-conda
2. In the same conda prompt or terminal where you installed your Cellpose, type:
   python -m cellpose --Zstack
   **You need this step to use the graphical user interface (GUI) to make sure Cellpose reads the stack correctly. You can try the default GUI, but there might be instances that the image is not formatted correctly and it will open it as a 2D image.
   https://cellpose.readthedocs.io/en/latest/do3d.html#input-format
3. Once the GUI is open, you can drag and drop your stack of images as a single file or use File > Load image > Select your image.
4. We recommend the Graphical User Interface (GUI) for the initial exploration of parameters that fit the most.
   **For this initial exploration, we are looking for the least time-consuming experience whilst obtaining the best segmentation possible. Here, a good cell segmentation means that as many cells as possible are visually represented correctly: cells occupy all their cytoplasm space including the edge membrane without any empty space; the same cell is identified correctly throughout the stack; cells are not fragmented into different objects (or identifiers/colours).
5. Calibrate automatically the cell diameter by pressing 'calibrate' under 'Segmentation' on the left side panel. The number next to the button may change from 30 to the auto calibrated. You can adjust it further based on the segmentation output.
   **Images are rescaled to match the images the model was trained from.
   **If cells are looking fragmented, it means you might need to increase the cell diameter. If cells are merged together, it might mean you should decrease this parameter.

6. Calibrate the 'Z-aspect'. You can view the orthogonal view to see if cells have their right shape. If not, change it by dividing the resolution on the z-axis by the resolution on the x or y-axis.
7. Obtain a segmentation using the pre-trained model 'cyto3'. Click on the button 'run cyto3'. Estimated time depends on the size of the image:
https://cellpose.readthedocs.io/en/latest/benchmark.html
** Depending on your images and markers used you may want to use a different model (please, check https://cellpose.readthedocs.io/en/latest/models.html):
    a. Cyto3. This model has been trained on images with cells (single or epithelial cells) that have a cytoplasmic label. **Recommended for epithelial cells**
    b. Nuclei. The nuclear model was trained using nuclei images, and, thus, we recommend it for round cells or nuclei.
    c. Tissuenet. This model was trained with images from tissues.
        **Unexpectedly, it performed worse than cyto3' for our images.
    d. Others. There are more models (yeast, bacteria, …), but there seems to be a consensus that the more general ones, typically, work better.

We recommend trying different ones and seeing which one gives you the best visual outcome. For our epithelial images, 'cyto3' is the model that has given us the best results. We also found an improvement from 'cyto2' to 'cyto3'. Thus, we encourage people to try the latest model available.

8. You can move up and down in the z-axis using the bottom scrollbar to inspect how good the segmentation is. Our best parameters were:
    - Model: 'cyto3'
    - Cell diameter: 60
      ** Depending on different zoom values, this value might vary.
    - Stitch threshold: 0.05
    Additional settings:
    - Flow threshold: -
      **Note that this parameter is not being used in 3D.
      https://cellpose.readthedocs.io/en/latest/do3d.html#segmentation-settings
    - Cell probability threshold: 0 or lower
      ** Negative numbers have also given us good outcomes, which helped to reach the membrane and fill all the space.
9. Even with the best parameters, there will be mistakes to be fixed. In 3D, typical mistakes will be cells that are not correctly tracked in the Z-axis. For this purpose, we perform manual corrections, explained in the next section and an additional stitching step (section 4.4).

**The code to reproduce these steps can be found in the '3D-deep-segmentation-protocol' jupyter notebook (see section "1. Initial segmentation: Cellpose").
**More information about Cellpose: https://cellpose.readthedocs.io/en/latest/index.html

### 4.2. Manual correction of labels

1. Install napari (16) in a new environment following these instructions:
https://napari.org/stable/tutorials/fundamentals/installation.html#napari-installation
2. Open napari by typing 'napari' in the terminal window. Install the following plugins: "devbio-napari" through the user interface under Plugins > Install/Uninstall plugins

\*\*This plugin is a compilation of very useful packages; you can look for relevant plugins on the napari hub https://www.napari-hub.org
\*\* Note that the first time napari is opened it might take a couple of minutes to launch.
3. Drag and drop the raw file fluorescence image, and the segmented image to be corrected into the napari window.
4. Right-click on the segmented image > Convert to labels (in case it is not already loaded as a labels layer). Select that layer by clicking on it. Decrease its opacity to see better the 'raw' image (recommended value of 0.25). Enhance the contrast of the raw image until you see the edges of the cells, by sliding the left side button towards the left side.
\*\*You should be able to see both the labels image and the cells' real edges.
\*\*You can also quickly toggle the visibility of a layer to check if the segmentation matches the cells' shape.
5. Manually correct the cells using the following tools:
    a. Dropper or 'Pick model'. Use it to identify the ID of the cell. When you click on a cell, the label will change to a number.
    b. 'Shuffle colours'. Use it to check if cells represented by the same colour are a single cell, which you would have to split into two, or different ones.
    \*\*Note that a given colour might be representing two different cells due to the reduced number of colours in the palette. Using 'Shuffle colours' will help you with that.
    c. 'Paintbrush'. To edit the segmented labels, simply pick a cell's ID to paint. Use the brush to modify regions belonging to that cell.
    d. 'Fill bucket'. Use it to fill empty (background) spaces with the selected cell ID.
    e. 'Eraser'. Remove regions of segmented areas transforming them into background.
    f. 'n edit dim'. You can use the 'Paintbrush', 'Fill bucket', and 'Eraser' in 2D or 3D. If 'n edit dim' is 2, these tools will only affect the selected z-slice. If it is 3, you will be editing in three dimensions (layers above or below, based on the brush size).
    g. 'contiguous'. (only 'Fill bucket') If checked, it will only change pixels whose values are the same as the selected one and connected to it. If not checked, and if (for instance) you pick the background ID, all the background will be filled.
    h. 'preserve labels'. If checked, only the background pixels will be modified by the 'Paintbrush', 'Fill Bucket', and 'Eraser'.
    i. 'Show selected'. It will only show the selected cell 'ID'.
    \*\*Very useful to see if there are fragmented cells.
    j. 'Plugins > napari-segment-blobs-and-things-with-membranes > Manually merge labels'.
        i. Create a 'New points layer' by clicking on the button most to the left.
        ii. Select 'Add point' on the top of your left bar.
        iii. Click in the centre of your cells to put them together as one. Press 'run'.
        \*\*You should see how the cells are now the same.
    \*\*Delete the points layer afterwards by clicking on the bin.
    k. 'Plugins > napari-segment-blobs-and-things-with-membranes > Manually split labels'.
        i. Create a 'New points layer' by clicking on the button most to the left.
        ii. Select 'Add point' on the top of your left bar.
        iii. Click in the centre of your cells to be split. Press 'run'.

**You can do this process with multiple cells.
		**You might need to correct the splitting in 3D.
		**Both functions perform the split and merging in 3D. If you just want to perform in 2D for a single slice, we recommend using the brush.
    6. Time of corrections depends on the quality of the segmentation obtained after automatic corrections and the quality of the raw image.
    ** For us, on average, a week of a person working 37.5 hours per week on a full stack with 100 cells.

** We found that it's best to first manually correct labels on each 2D slice, then perform 3D automatic corrections with TrackMate. If, for example, there are cells fused together or split into multiple parts in 2D, the 3D stitching in TrackMate will be impacted. Thus, we recommend first having a good 2D segmentation and then, focusing on improving its 3D stitching (see section 4.4).

To reproduce this section, check this section from the following jupyter notebook here.

### 4.3. Segmentation quality assessment

For heterogeneous cell shapes along the z axis, we found that most manual correction was needed to improve the 3D stitching rather than the 2D x-y planes. In addition to visually assessing the quality of segmentation, we introduce a new metric to evaluate the 3D stitching quality: Cell Persistence score, which assesses how many labels (cells) are followed through a minimum % of z planes. A perfect Cell Persistence score would be achieved when each cell is tracked throughout the whole stack with no interruption.
  1. Use your previously set up napari environment to install the plugin 'EpiTools' https://github.com/epitools/epitools.
  2. To calculate the Cell Persistence score, click on 'Plugins > Calculate Cell Persistence score'. A widget will open on the right-hand side of the GUI.
  3. Select your raw image in 'image', and your segmented image in 'labels'.
     a. The percentage of z-slices indicate the number of z-planes required for cells to be counted as successfully stitched in 3D.
        ** threshold values can be adjusted depending on how strict users want the stitching to be, for example we used 80%.
     b. Show overlay. If selected, it will create another labels layer with the cells that fit the criteria.
     c. Run metrics. If ticked, it will run our module to calculate cell statistics on only the corrected cells.
  4. Press 'Run'. After a while, it will output the number of good cells that have been successfully segmented continuously in the selected percentage of z-planes.
     ** Use this function to overlay a selection of the best reconstructed cells in 3D, to help identify problem areas that need further correcting. If a patch of successfully reconstructed cells is identified and has sufficient N number, the exported layer can be used for quantification ticking 'run metrics' or exported for further analysis.
** Alternatively, you can compute the Cell Persistence Score in a Python console or Google Colab.
This step is quite important and specific to pseudostratified or columnar tissues, where we expect most of the cells to be present in all layers from the top to the bottom of the tissue.

## 4.4. Tracking as custom z-stitching

For very tall tissues where cells change shape and position along z, we found that the built-in stitching in Cellpose was not sufficient to correctly follow and stitch the cells along z. We found that the tracking package TrackMate could be used as a powerful custom z-stitching algorithm. The main advantage is the possibility of setting multiple custom parameters, such as the number of planes an object is allowed to "disappear" and the distance that the object can travel between frames.

1. Download FIJI (Fiji is just ImageJ) from https://imagej.net/software/fiji/downloads. It is a portable application, so you only need to unpack and open it by double-clicking on the FIJI executable.
2. Open your 3D segmented image to be corrected.
3. Click on Plugins > Tracking > TrackMate. There would be a pop-window asking you to swap the z and time-axis. The answer should be 'Yes'.
   \*\*If you do not swap these axes, you would see circles around the cells instead of the boundaries of each cell in pink.
4. Click 'next'. Select the 'Label image detector' which is providing the segmented file and click 'Next'.
   \*\*You can preview to check if the IDs are ok.
5. It will process the labels. Once it is finished, click 'Next'.
6. Do not change the initial threshold, click 'Next'. It will process it again, click 'Next'.
7. Select a 'tracker', we used 'Kalman tracker'. 'Next'.
8. Set the tracker parameters: we found that slightly increasing the search radius compared to default parameters improved the tracking: set 'Initial radius' to 20, 'Search radius' to 25 and 'Number of missing frames' to 2. Click 'Next'. Once it is processed, click 'Next' again and a second time.
9. In the 'Set filters on tracks' window, pick relevant tracking metrics (e.g., distance travelled, splitting events) to apply a colour coding and evaluate the tracking. If needed, incorrect tracks can be filtered out at this stage.
10. Check what you want to display and click 'Next', and 'Next' again.
11. Export your new labelled image by picking 'Export label image' and clicking 'Execute'. Leave it as is and click 'OK'.
12. You can now save the new segmented image.
    \*\*Note that IDs might differ from the original IDs. Thus, the cell with (e.g.) ID 1 in the original image might not be the cell with ID 1 in the new segmented image.

## 4.5. Fine-tuning initial segmentation

1. Cellpose pre-trained models were trained using 2D images. Therefore, to re-train any of the available models, 2D images should be used as input. To do that, 3D images need to be transformed into XY, YZ, and XZ slices.
    a. We have developed a code to generate 3D sections (section 5 "Refining the segmentation: Cellpose fine-tuning").
    b. Code from Cellpose is also now available to do this.
       https://github.com/MouseLand/cellpose/blob/main/cellpose/gui/make_train.py

2. Split the dataset into training and test sets either manually or with a python function like 'train_test_split' from Scikit-learn (by default 25% test, 75% training). This is a step required to train your data.
3. Re-train a given model with your ground truth data (transfer learning). We recommend keeping the default parameters.
    \*\*Based on our quantifications, depending on the number of images used, more epochs will give a better prediction, but it takes longer (3468 2D slices images, 2000 epochs: 151393s seconds; 100 epochs: 7757s seconds).
    \*\*Estimated time: dependent on number of 2D images; for 3468 2D images, approx. 2 days of computing time with the workstations described above). For additional information, please check:
    https://cellpose.readthedocs.io/en/latest/benchmark.html
    Additionally, we encountered some common issues with this step and provide possible solutions:
    a. Problem with the minimum number of labels on an image.
        \*\*Fix: use '--min_train_masks' with 0.
    b. Overfitting issues. If you re-train a model with images with all cells and no background, when you input images with some empty space (background) that space will be filled with cells.
        \*\*Fix: use images with some black background or empty space (no labels) in your images.
    c. Even though we recommend to re-train your model using python, you can also use the GUI to re-train it. However, you might get: 'IndexError: list index out of range'.
        \*\*Fix: you would need to obtain the _seg.npy files beforehand in python. \*\*Note: this error will only happen using the GUI of cellpose.
4. Predict images with your new model.
    a. Click Models > Add custom torch model to GUI > Pick the model file you have created.
    b. Under 'Other models', you should find your new model by clicking on the drop-down button 'custom models'. Pick your model.
    c. Click 'run'.

\*\*This code can be reproduced at '3D-deep-segmentation-protocol' jupyter notebook (section "Refining the segmentation: Cellpose fine-tuning").

## 5. 3D Analysis and visualisation with napari

### 5.1. Visualization

1. Use the same environment you used for the 'manual segmentation' (see "Manual segmentation" section).
2. Open napari and drag and drop your raw image.
3. Drag and drop your segmented image. If it is not seen as a 'labelled image', right-click on the layer corresponding to the segmentation and select 'convert to labels'.
    \*\*If you do not see your labels layer, it could be hidden behind another layer. The first image in the list is the top layer and layers below will be only visible if opacity allows.

4. To visualise the cells in 3D, click on the button (2nd starting from the left, left-bottom panel) to swap from 2D to 3D view.
   \*\*You may see that your cells do not look as expected (e.g., much flatter). If so, you may have to update the voxel size to reflect the z-slicing of the data.
5. To change the voxel size of both your images: 'Tools > Utilities > Set voxel size of all layers'. Change the values to be like your acquired image properties. \*\*We typically open an image, 'FIJI > Image > Properties' and copy the voxel height, width and depth to napari.
   \*\*Because you have changed the view, your image may have disappeared. Reset to its original view (house icon, left-bottom panel) and then, click on the button to change to 3D view.
6. Click 'Run'. You should see your cells with the correct aspect ratio.
   \*\*We recommend visualising your raw image with the rendering engine "Attenuated_mip", where you can adjust the attenuation depending on your images. If you want crispier images, you may change its interpolation.
7. You can now export your cells as displayed by clicking on 'File > Save screenshot…'
8. (Optional) To save a 3D animated video of your 3D stack, you can use the napari-animation plugin https://github.com/napari/napari-animation.

### 5.2. Analysis

1. To obtain cellular features, we recommend using the 'regionprops' function in napari: 'Tools > Measurements table > Regionpops'. It will appear on the top-right as a new panel with the settings.
2. Make sure your raw image is selected on 'image' and segmented on 'labels'.
3. Tick or untick the features of interest and click on 'Run'.
   \*\*Depending on your question of study you might pick different measurements, but we recommend to tick: 'size', 'intensity', 'perimeter', and 'shape'.
4. On the bottom side you will see a table with each individual segmented cell as a row and the columns as features.
   \*\*Note that the columns are named with the same name as in 2D, but they are 3D features. Thus, 'area' would be 'volume', and so on.
   https://github.com/haesleinhuepf/napari-skimage-regionprops
5. To save the table you can either copy to the clipboard (clicking on that button) and paste it on any processing software sheet (e.g. Microsoft Excel), or you can save it as a comma-separated-values (or csv) file.
   \*\*You can use this table to search for possible outliers, like cells with very small/large volumes or weird shapes. You can display cells that look odd by using the 'show selected' feature (as explained before) and correct it if necessary.

# Discussion

Here we have detailed an end-to-end protocol covering all aspects from sample preparation and imaging to 3D segmentation and quantification, which has been optimised to analyse individual cells in *Drosophila* wing discs. Improvements in sample mounting and imaging have been essential for our ability to segment epithelial cells individually within a 50 µm thick tissue.

The largest enhancements have been using 2-photon for excitation and imaging the wing discs with a water dipping objective, which minimizes refractive index mismatches between the sample and the objective. For this, we describe an optimised sample mounting method that leverages on a biocompatible adhesive to affix wing discs to the bottom of a cell culture dish, filled with growth media. This mounting method is suitable for a wide variety of tissues and tissue slices.

In terms of the image analysis aspects, we presented a human-in-the-loop pipeline that leverages on existing open-source software. In terms of pre-processing, we tested two image restoration method and found that, unexpectedly, they did not aid in cell segmentation even if the visual appearance of images improved. In particular, Cellpose models appeared to perform worse on deconvolved data, probably because they were trained on raw, noisy images. For segmentation, we find that Cellpose pre-trained models work reasonably well 'out-of-the-box'. Crucially, the quality of the segmentation substantially increases if we fine-tune the model using our dataset, even starting from a single image stack containing hundreds of cells. While we acknowledge the powerful properties of fine-tuning, we also want to highlight that it can be counterproductive. As explained in section 4.4, if a model is fine-tuned with images completely filled by cells (no background areas), the newly trained model might always predict cells, even in background regions, an indication of model overfitting.

In complex pseudostratified tissues like the *Drosophila* wing disc, a main challenge is stitching of the 2D segmentations along the z-axis and we found that commonly used quality metrics (e.g. intersection over union (21)) were not useful to assess this when ground truth data is not available. Thus, we introduced the Cell Persistence Score, which quantifies how well the cells are being tracked in the Z axis. We have found that using the Cell Persistence score aids us to accurately assess how the segmentation improves during manual correction and to identify problematic regions in the image.

We expect our protocol to shed light into the challenging task of analysing thick tissues at a cellular level, offering a robust pipeline adaptable to other complex biological systems. In the future, aided by advancements in computational tools and imaging techniques, we expect to see more quantitative analysis of tissues across a wide range of biological contexts.

# Acknowledgments

G.P. was supported by an EMBO Long-Term Fellowship (ALTF 786-2020). P.V.-M. was supported by EPSRC grant EP/X03139X/1. R.B. was funded by a UCL COMPLEX PhD studentship, and a Leverhulme Trust Project Grant RPG-2020-068 awarded to Y.M. Y.M. was supported by the MRC award MR/W027437/1, a Lister Institute Research Prize and EMBO Young Investigator Programme.

# Author contributions

G.P. performed and supervised wet lab experiments and analysed data. P.V.-M. designed the segmentation protocol. I.F.-M. acquired and annotated Drosophila wing disc. A.M. worked on the segmentation analysis and obtained new models. K.L acquired and annotated Drosophila wing disc. Q.Z. improved the segmentation protocol using TrackMate. R.B. helped optimise sample preparation and imaging. Y.M. conceived the idea and supervised the project. G.P, P.V.-M. and Y.M. wrote the manuscript.

**Competing interests**

There are no competing interests to declare.

# References

1. Hampson KM, Turcotte R, Miller DT, Kurokawa K, Males JR, Ji N, et al. Adaptive optics for high-resolution imaging. Nat Rev Methods Primers. 2021;1.
2. Luu P, Fraser SE, Schneider F. More than double the fun with two-photon excitation microscopy. Commun Biol. 2024;7(1):364.
3. Greener JG, Kandathil SM, Moffat L, Jones DT. A guide to machine learning for biologists. Nat Rev Mol Cell Biol. 2022;23(1):40-55.
4. Mahmud M, Kaiser MS, McGinnity TM, Hussain A. Deep Learning in Mining Biological Data. Cognit Comput. 2021;13(1):1-33.
5. Hallou A, Yevick HG, Dumitrascu B, Uhlmann V. Deep learning for bioimage analysis in developmental biology. Development. 2021;148(18).
6. Andres-San Roman JA, Gordillo-Vazquez C, Franco-Barranco D, Morato L, Fernandez-Espartero CH, Baonza G, et al. CartoCell, a high-content pipeline for 3D image analysis, unveils cell morphology patterns in epithelia. Cell Rep Methods. 2023;3(10):100597.
7. Kar A, Petit M, Refahi Y, Cerutti G, Godin C, Traas J. Benchmarking of deep learning algorithms for 3D instance segmentation of confocal image datasets. PLoS Comput Biol. 2022;18(4):e1009879.
8. Wolny A, Cerrone L, Vijayan A, Tofanelli R, Barro AV, Louveaux M, et al. Accurate and versatile 3D segmentation of plant tissues at cellular resolution. Elife. 2020;9.
9. Pachitariu M, Stringer C. Cellpose 2.0: how to train your own model. Nat Methods. 2022;19(12):1634-41.
10. von Chamier L, Laine RF, Jukkala J, Spahn C, Krentzel D, Nehme E, et al. Democratising deep learning for microscopy with ZeroCostDL4Mic. Nat Commun. 2021;12(1):2276.
11. Nematbakhsh A, Levis M, Kumar N, Chen W, Zartman JJ, Alber M. Epithelial organ shape is generated by patterned actomyosin contractility and maintained by the extracellular matrix. PLoS Comput Biol. 2020;16(8):e1008105.

12.	Gomez-Galvez P, Vicente-Munuera P, Tagua A, Forja C, Castro AM, Letran M, et al. Scutoids are a geometrical solution to three-dimensional packing of epithelia. Nat Commun. 2018;9(1):2960.
13.	Kluyver T, Ragan-Kelley B, Pérez F, Granger B, Bussonnier M, Frederic J, et al. Jupyter Notebooks–a publishing format for reproducible computational workflows.  Positioning and power in academic publishing: Players, agents and agendas: IOS press; 2016. p. 87-90.
14.	Stringer C, Pachitariu M. Cellpose3: one-click image restoration for improved cellular segmentation. bioRxiv. 2024:2024.02.10.579780.
15.	Tinevez JY, Perry N, Schindelin J, Hoopes GM, Reynolds GD, Laplantine E, et al. TrackMate: An open and extensible platform for single-particle tracking. Methods. 2017;115:80-90.
16.	contributors n. napari: a multi-dimensional image viewer for python. 2019.
17.	Stringer C, Wang T, Michaelos M, Pachitariu M. Cellpose: a generalist algorithm for cellular segmentation. Nat Methods. 2021;18(1):100-6.
18.	Heller D, Hoppe A, Restrepo S, Gatti L, Tournier AL, Tapon N, et al. EpiTools: An Open-Source Image Analysis Toolkit for Quantifying Epithelial Growth Dynamics. Dev Cell. 2016;36(1):103-16.
19.	Roddy P, Matthews, D., J Smith, P., Felder, A., Vicente Munuera, P., Paci, G., & Mao, Y. . Epitools (Version 0.0.12) 2023.
20.	Schindelin J, Arganda-Carreras I, Frise E, Kaynig V, Longair M, Pietzsch T, et al. Fiji: an open-source platform for biological-image analysis. Nat Methods. 2012;9(7):676-82.
21.	Hirling D, Tasnadi E, Caicedo J, Caroprese MV, Sjogren R, Aubreville M, et al. Segmentation metric misinterpretations in bioimage analysis. Nat Methods. 2024;21(2):213-6.
22.	Berg S, Kutra D, Kroeger T, Straehle CN, Kausler BX, Haubold C, et al. ilastik: interactive machine learning for (bio)image analysis. Nat Methods. 2019;16(12):1226-32.